\documentclass[preprint,12pt]{elsarticle}

\usepackage{amssymb}
\usepackage{amsmath,epsfig}

\journal{Acta Astronautica}

\begin{document}

\begin{frontmatter}



\title{SETI: The transmission rate of radio communication and the  signal's detection}

\author{P. A. Fridman}

\address{ASTRON, Oude Hoogevensedijk 4, 7991PD Dwingeloo, The Netherlands}

\begin{abstract}
The transmission rate of communication between  radio telescopes on Earth and extraterrestrial intelligence (ETI) is here calculated  up to  distances of 1000 light years. Both phase-shift-keying (PSK)  and frequency-shift keying (FSK)  modulation schemes are considered. It is shown that M-ary FSK is  advantageous in terms of energy. Narrow-band pulses scattered over the spectrum sharing a common drift rate can be the probable signals of ETI. Modern SETI spectrum analyzers are well suited to searching for these types of signals. Such signals can be detected using the Hough transform which is a dedicated tool for detecting patterns in an image.  The time-frequency plane representing the power output of the spectrum analyzer during the search for ETI gives an image from which the Hough transform (HT) can detect signal patterns with frequency drift.

\end{abstract}

\begin{keyword}
SETI; communication; transmission rate; Hough transform

\end{keyword}

\end{frontmatter}


\section{Introduction}
\label{int}

 Up until the present day searches have been undertaken for signals of two types of extra-terrestrial intelligence (ETI): monochromatic (continuous wave, CW) and pulses, \citep{tart1}, \citep{seti}.
  The detection of any such signal is presumed to be a demonstration
of the existence of ETI. The signals are expected from both planetary systems
similar to our Solar system and from beacons scattered across our Galaxy.

 During the search for extraterrestrial intelligence (SETI), both targeted searches  and sky surveys   have been performed to detect   narrow-band or pulse-like signals, \citep{seti}.
The signals are examined by   multi-channels spectrum analyzers. There can be millions of such channels and the bandwidth of each channel is of 1 Herz or less. This approach was proposed in the Project Cyclops report \citep{cycl} and in a NASA multi-channel spectrum analyzer article \citep{cull}. More recent examples of these analyzers have been described in projects  Phoenix \citep{phoenix}, (1 Hz between 1,000 and 3,000 MHz), SERENDIP IV \citep{serendipIV}, (168Mchannels, total bandwidth 100MHz), SERENDIP V \citep{serendipV}, (128Mchannels, total bandwidth 200MHz).

 Many proposals have been made to detect leakage emissions which manifest  radio activity of  technologically developed civilizations. An estimate of the possibilities of such eavesdropping are given in \citep{tart}.

It is reasonable to assume that an  ETI  sends signals  not simply to attract attention, but to deliver some useful information, i.e., an ETI sends  some modulated signals containing messages.  In each epoch  we can only use our current technical knowledge to predict what kind of signals an ETI could send.  This paper is dedicated to the estimation of the communication transmission rate with ETI in a sphere of radius 1000 light years (ly) which contains $\approx 2.3 \cdot 10^{6}$ F0 through K7 stars.

The two primary resources for a  communication system are  transmitter power and  channel bandwidth. These two parameters are usually the subject of a tradeoff: one being more precious than the other one. Accordingly,  communication systems are classified as being either power-limited or bandwidth-limited. In power-limited systems, coding schemes are used to save power at the expense of  bandwidth, whereas in band-limited systems spectrally efficient modulation techniques can be used to save the bandwidth at the expense of the power. In this paper it is assumed    that, in the case of SETI, the transmission power is a more precious parameter than the transmission bandwidth,  hence  the ETI communication system should be classified as {\it power-limited}.\\
Un-coded  modulation methods are considered: phase-shift keying (PSK) and frequency-shift keying (FSK). Pulse-amplitude modulation (PAM) has inferior characteristics to PSK and FSK and the estimates for PAM are not given here.

Having some plausible estimates of the transmission rate for different modulation schemes we can consider an ``inverse''  SETI approach:  assuming that  an ETI uses   certain signals (which our civilization would use), for the transmission of information, the  pragmatical solution  would be to search for just these kinds of signals   using appropriate detection algorithms.

\section{The transmission rate}

\subsection{Radio link power budget}

The communication transmission rate is calculated here under the following assumptions:

1. The distance from Earth to an ETI  transmitter is $R<1000ly$.

2. The communication is conducted at S- and X-bands.

3. The transmitter power of ETI is $P_{tr}=10^{12}W$.

4. The transmitting antenna is omni-directional, i.e., the antenna gain is $G_{tr}=1$ and effective radiated power (EIRP) is equal to $10^{12}W$.

5. The Arecibo radio telescope has been chosen as the receiver  complex: the illuminated diameter of the dish is $d=225m$, antenna/feed efficiency $\eta=0.74$, system temperature $T_{\rm sys}=32$~K which gives the {\it system equivalent flux density} (SEFD) $\approx 3J$.

\subsection{Bit error and transmission rate}

The signal-to-noise ratio for the received signal in the bandwidth
equal to 1~Hz is
\begin{equation}
SNR_{\rm power}=\frac{P_{tr}G_{tr}A_{\rm eff}}{4\pi R^{2}k_{B}T_{\rm
sys}},\text{ \ \ \ }k_{B}=1.38\times 10^{-23}\,\, {\rm J/K}.
\end{equation}

The potential transmission bit rate can be estimated using the
Shannon's formula for channel information throughput with the
bandwidth $\Delta F$ and a given $SNR$ as:
\begin{equation}
C =\Delta F\log _{2}(1+\frac{SNR_{\rm power}}{\Delta F}).
\end{equation}
This formula sets a limit on the transmission rate, but not on the error probability. Since the latter is an important characteristic of a communication system, rather than $SNR_{\rm power}$, another figure of merit is widely used:
\begin{equation}
SNR_{\rm bit}= SNR_{\rm power}T_{b}=\frac{SNR_{\rm
power}}{RW},
\end{equation}
where $SNR_{\rm bit}$ is the signal-to-noise ratio per bit, $T_{\rm b}$ is the
duration of signal transmission per 1 bit and $RW$ is the
transmission rate. Using this new notation and making the transmission bit rate equal to the channel capacity, $RW=C$, we have:
\begin{equation}
C=\Delta F\log _{2}(1+SNR_{\rm bit}\frac{C}{\Delta F})
\end{equation}
and
\begin{equation}
SNR_{\rm bit}=\frac{\Delta F}{C}(2^{C/\Delta F}-1)
\end{equation}
 If $SNR_{\rm bit}$ becomes small, the ratio $(C/\Delta F)\rightarrow 0$ and the corresponding value $SNR_{\rm bit}=\frac{1}{\log _{2}e}=0.693$ or, in decibels, $SNR_{\rm bit}= -1.6$~dB. This value of $SNR_{\rm bit}$ is called the {\it Shannon limit}.

\subsection{PSK}
 A digital modulation scheme with phase-shift keying (PSK) is
widely used in space communication systems \citep{yuen}. It provides both a low
bit-error-rate (BER) and minimum bandwidth.
BER for the binary phase-shift keying (BPSK) and for  other modulation methods  will be calculated using the formulas in Appendix A.

Quadri-phase shift keying (M-ary phase modulation, $M=4$), denoted as QPSK, has the same bit-error performance as BPSK, but the equivalent bit rate is two times higher than that of BPSK.

   To obtain $BER=10^{-5}$ for un-coded BPSK the necessary  $SNR_{bit,BER=10^{-5}}$ must be equal to 9.6 dB (or 9.0945). Therefore, the ratio
$SNR_{power}/SNR_{bit,BER=10^{-5}}$ gives the transmission rate achievable for a given $SNR_{power}$ as a function  of $P_{tr}, G_{tr}, R, A_{eff}$ for the
value $BER=10^{-5}$.

\begin{figure}[h]

\epsfig{figure=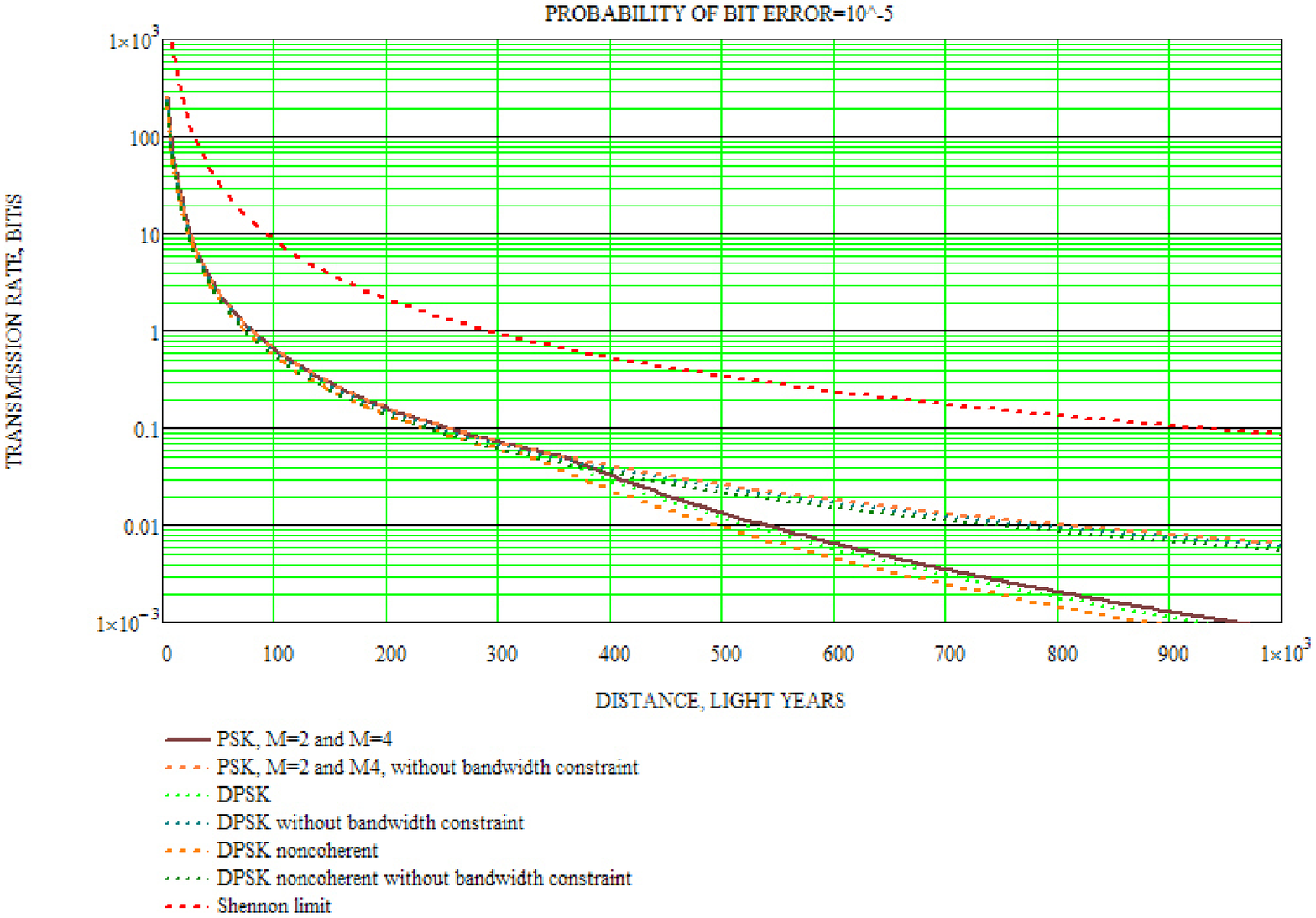,height=19.0cm,width=17.0cm}
\caption{The transmission rate as a function of distance for un-coded PSK, transmitter $EIRP=10^{12}W$ (transmitting power=$10^{12}W$ and omnidirectional antenna) and Arecibo-like receiver antenna, $SEFD=3J$, $BER=10^{-5}$.  The top curve shows the Shannon limit (best possible rate). Breakdowns in two collections of curves (PSK, DPSK and noncoherent DPSK) are produced by  bandwidth constraint due to  interstellar propagation at $\Delta f_{min}=0.05Hz$.  }
\label{f1}
\end{figure}

 Fig. \ref{f1} demonstrates how the transmission rate depends upon the  distance (in light years) for the Arecibo radio telescope and  $BER=10^{-5}$,  $ P_{tr}=10^{12}~W,
G_{tr}=1$. The upper curve corresponds to the Shannon limit. The  three closely running curves below correspond to PSK, coherent and non-coherent differential PSK (DPSK), respectively. For the coherent DPSK  $SNR_{bit,BER=10^{-5}}=9.76$ and for the non-coherent DPSK $SNR_{bit,BER=10^{-5}}=10.82$, \citep{Okunev}.
There is no significant difference between these three curves, i.e.,  DPSK is only slightly inferior to PSK. DPSK does not require an elaborate method for estimating the carrier phase and it is often used in digital communication systems.

 Table \ref{TablePSK}  contains the values of transmission rate corresponding to these three curves.
\begin{table} [h]

    \caption[]{The transmission rate bit/sec as a function of distance for un-coded  PSK, transmitter $EIRP=10^{12}W$ and Arecibo-like receiver antenna, $SEFD=3J$, $BER=10^{-5}$. }

 \centering
    \label{TablePSK}
   \begin{tabular}{c|c|c|c|c|c}
    \hline

     \multicolumn{6}{c}{\hspace{3.0cm} Distance, light years}\\
     \hline
    PSK  & 10 & 50 &  100 & 500 &  1000 \\
     \hline
   coherent BPSK, & 64.7 & 2.56 &  0.65 & 0.026 &  $6.5 \cdot 10^{-3}$ \\
   coherent DPSK & 60.3 & 2.4 & 0.60 & 0.024 & $6.0 \cdot 10^{-3}$ \\
   non-coherent DPSK & 54.4 & 2.17 & 0.544 & 0.022 & $5.4 \cdot 10^{-3}$ \\
   \end{tabular}

\end{table}

The lower collection of curves (PSK and two DPSK) corresponds to the more realistic situation when scattering and multi-path caused by  the  interstellar medium is taken into account, \citep{cordes}. This effect  broadens the sinusoidal  signal up to the bandwidth $df_{min} = 0.01$ to $ 0.05 Hz$. Therefore, the assumption for the average interval  $T_{average},   1/T_{average}=\Delta f_{average} \approx 1/T_{bit}$ is not valid after $\Delta f_{average}<df_{min}$. Limiting the bandwidth by the value $df_{min}$ worsens the $SNR_{bit}=SNR_{\rm power}T_{b}/\sqrt{df_{min}T_{b}}=SNR_{\rm power}\sqrt{T_{b}/df_{min}}$. Therefore, the transmission rate for
$df_{min}>1/T_{b}=RW$ must be calculated as follows:
\begin{equation}
RW=(\frac{SNR_{power}}{SNR_{bit,BER=10^{-5}}})^{2}\frac{1}{df_{min}}
\end{equation}

The breakdown  visible in Fig. \ref{f1} corresponds to  $df_{min}=0.05Hz$, i.e., in the case of very long ($> 20 s$) narrow-band signals used for communication the transmission rate is significantly lower than the potential one.
\begin{figure}[h]
\epsfig{figure=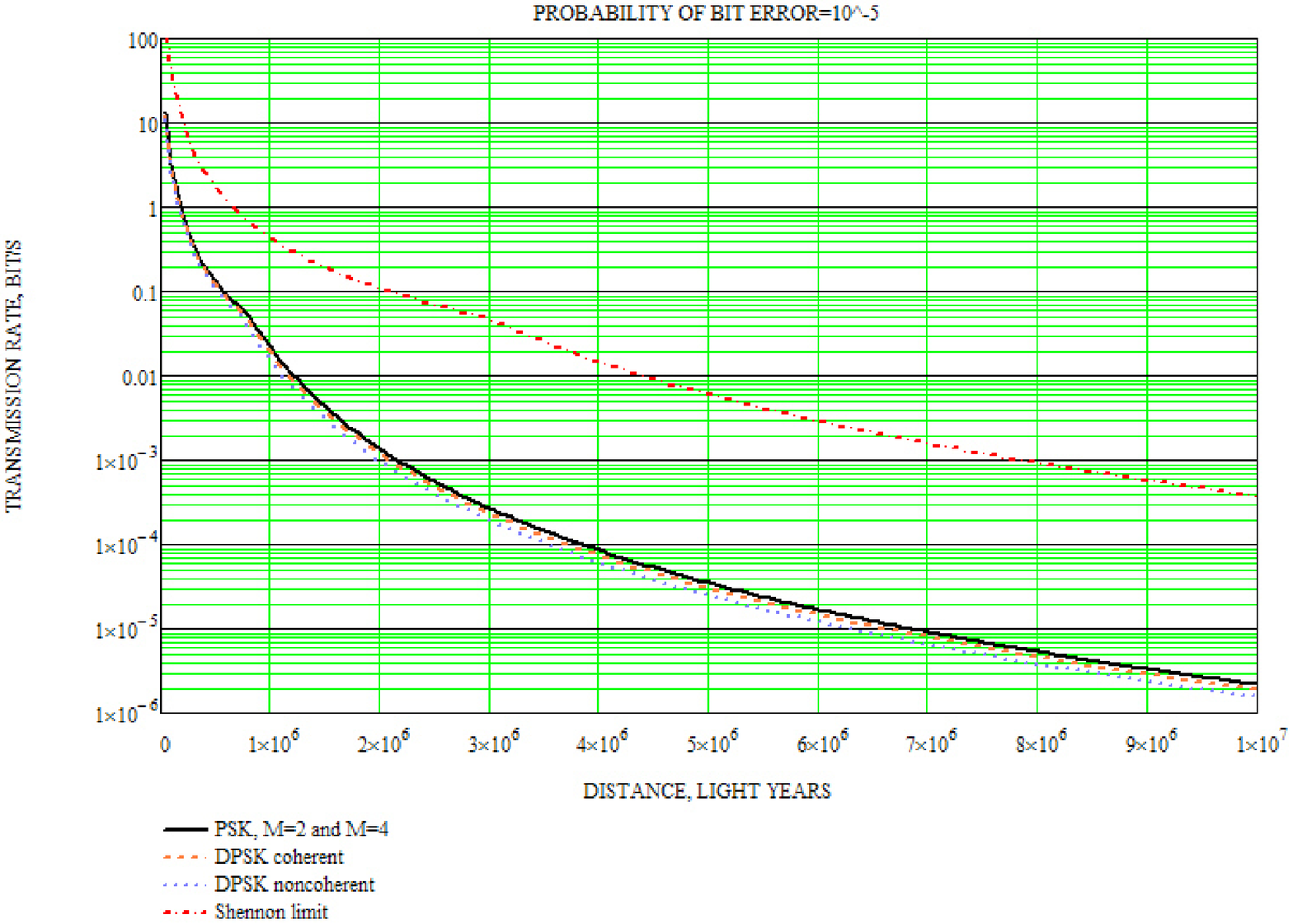,height=16.0cm,width=16.0cm}
\caption{The increase in transmission rate using a high gain Arecibo-like transmitting antenna. Other parameters as in Fig. 1. Note the change in scale on both axes. The bandwidth constraint due to the interstellar propagation is $\Delta f_{min}=0.05Hz$. }
\label{f2}
\end{figure}

Fig. \ref{f1} depicts the situation with the omni-directional transmitting antenna. When both transmitter and receiver sides know each other's coordinates (targeted communication), a directional transmitting antenna similar to the receiver antenna (Arecibo in our example) can significantly increase the communication distance, see Fig. \ref{f2}. The large antenna gain $G_{tr}$ at 1GHz allows for communication  to several Mly.

A considerable gain in BER performance can be achieved using special coding of the bit stream before modulation. For example, for binary modulation and $P_{B}=10^{-5}$ the error performance of turbo code is within 1.8 dB of the Shannon limit. But it is difficult to guess the coding scheme used by ETI, therefore, the performance of communication is calculated here for un-coded signals only.

The first phase of SKA presumes $SEFD=2.7 J$ (250 dishes, each 15m diameter, antenna-feed efficiency $\eta=0.7, T_{sys}=30 K$), \citep{dewdney}. This is comparable with Arecibo, but the field-of-view is $1.37deg^{2}$ at 1GHz which in combination with multi-beaming  allows for a  faster all-sky survey.
The full SKA is to consist of 1500 dishes \citep{faulk}, but the core ($25\%$) will have $SEFD=1.785 J$ which will  increase the transmission rate by 1.68 times.

\subsection{FSK}

As was mentioned at the end of the Introduction the transmission power is a more precious parameter than is the transmission bandwidth in ETI communication. Using  of $M=2^{k}$ orthogonal signals can increase the necessary bandwidth but it requires smaller $SNR_{bit}$ to achieve a given error probability with FSK.  M-ary PSK provides better bits-per-second-per-hertz-of-bandwidth ratio but $SNR_{bit,BER=10^{-5}}$  for M=8, 16 and 32 is 19.8, 55.4 and 171.2, respectively. The same tendency exists for M-ary PAM (pulse amplitude modulation) and M-ary QAM (quadrature amplitude modulation).

The situation is different for  frequency shift keying (FSK) signals. Each symbol corresponding to a particular binary
sequence of the length $k$ is transmitted on one of the $m$-th
frequencies during an interval $T$:
\begin{equation}
s_{m}(t)=A\cos [2\pi (f_{c}+m\Delta f)t],0\leq
t\leq T,m=1,2,....M,
\end{equation}
where $\Delta f=1/2T, T-$ is the duration of a symbol. For
example, for $k=5$, there are $32$ different orthogonal signals,
each at its own frequency. The total  bandwidth is $W=M\Delta f$.
The effective transmission rate is $RW=k/T,\,\, k=log_{2}M$. The
ratio $RW/W$ is called {\it bandwidth expansion}:
$RW/W=(2log_{2}M)/M$. Theoretically, by increasing $M$ it is
possible to reach the Shannon limit.

  Un-coded M-ary frequency shift keying (MFSK) gives a higher transmission rate  than an un-coded BPSK for a fixed BER=$10^{-5}$ and
 $M\geq 8$. MFSK requires a wider bandwidth than  BPSK but this  resource  is of less importance than  transmitter power.

\begin{figure}[h]
\epsfig{figure=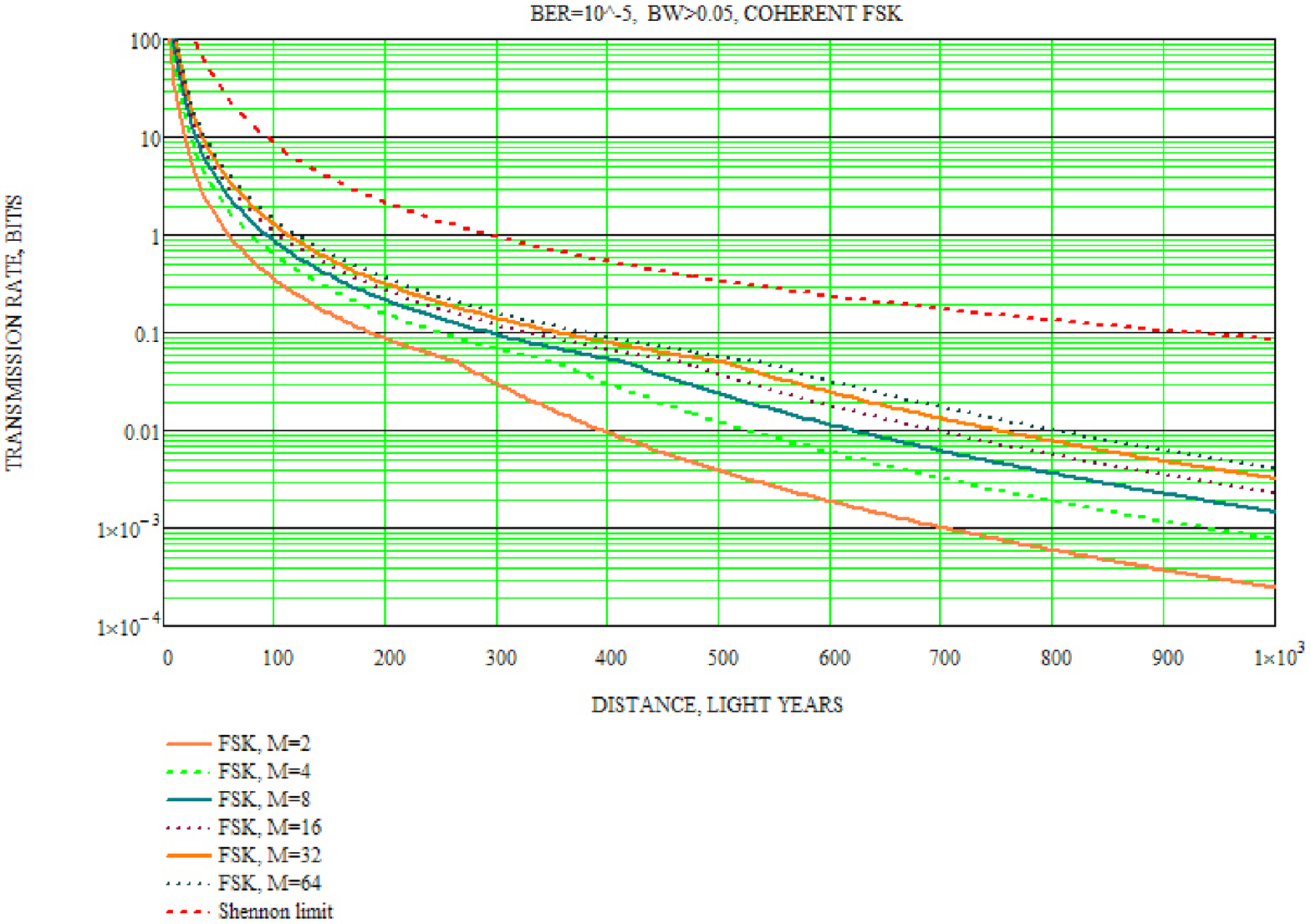,height=16.0cm,width=16.0cm}
\caption{ The transmission rate as a function of distance for un-coded coherent MFSK, transmitter $EIRP=10^{12}W$ and Arecibo-like receiver antenna, $SEFD=3J$, $BER=10^{-5}$. The bandwidth constraint due to the interstellar propagation is $\Delta f_{min}=0.05Hz$; note the resulting break at 0.05 bits/s.}
\label{f3}
\end{figure}

Fig. \ref{f3} shows the transmission rate as a function of distance for un-coded {\it coherent} MFSK, transmitter $EIRP=10^{12}W$ and Arecibo-like receiver antenna, $SEFD=3J$, $BER=10^{-5}$. Bandwidth constraint due to  interstellar propagation is $\Delta f_{min}=0.05Hz$. The Shennon limit is the upper curve. The lower curve corresponds to M=2 and the following curves correspond to M=4,8,16,32 and 64, respectively. The breakdown due to the loss of  effectiveness in averaging is clearly visible.

\begin{figure}[h]
\epsfig{figure=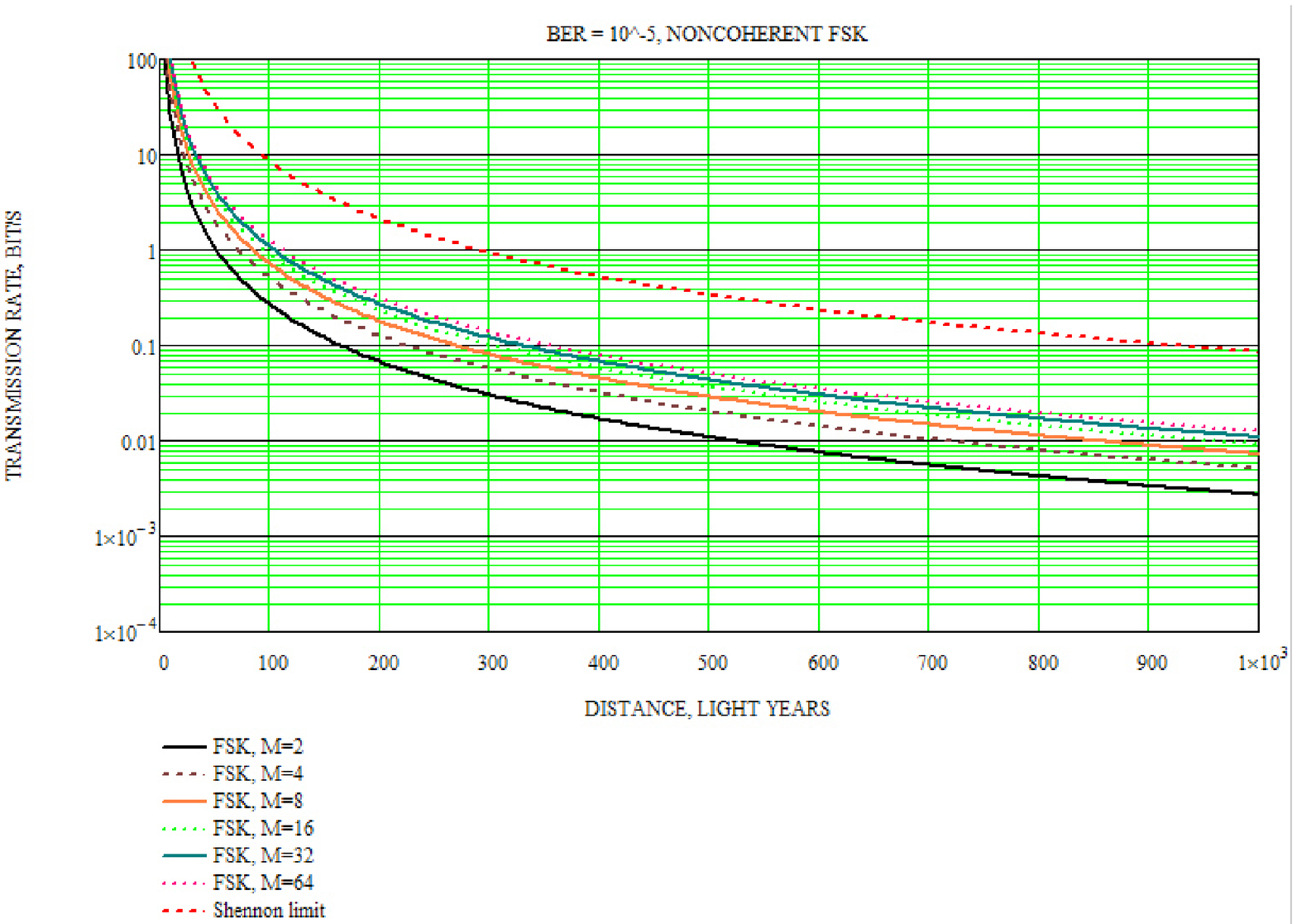,height=16.0cm,width=16.0cm}
\caption{The transmission rate as a function of distance for un-coded non-coherent MFSK, transmitter $EIRP=10^{12}W$ and Arecibo-like receiver antenna, $SEFD=3J$, $BER=10^{-5}$. Interstellar broadening effects are avoided by limiting pulse durations to less than 20 s, leading to higher transmission rates.}
\label{f4}
\end{figure}

All formulas used for calculation (see Appendix A) contain $SNR_{bit}$ or $SNR_{symbol}$ which, in their turn, depend on the {\it energy} spent during the transmission. Having a constant transmitter power $P_{tr}$, the larger the distance, the longer the necessary duration of the elementary signal. Another option is to keep  the energy per symbol constant. With this approach the duration $T$ of the signal can be shorter than $1/\Delta f_{min}$  whereas the $P_{tr}$ is larger. This transition to the shorter, pulse-like signals with a duration of less than $20$ to $100s$  means that  the breakdown due to $\Delta f_{min}$ can be avoided.\\
Fig. \ref{f4} shows  the transmission rate as a function of distance for un-coded {\it non-coherent} MFSK, the same Arecibo-like receiver antenna, $SEFD=3J$, $BER=10^{-5}$. The Shennon limit is represented by the upper curve, the lower curve corresponds to M=2 and the subsequent curves correspond to M=4,8,16,32 and 64, respectively. The transmission rate for non-coherent MFSK is slightly less than for the coherent MFSK, Fig. \ref{f3}. \\
There is no breakdown due to  $\Delta f_{min}$  in Fig. \ref{f4} deliberately in order to show the run of the curves in the absence of this effect and assuming that the transmitter keeps  the transmission energy per symbol constant, i.e., the duration of the symbol is limited by $1/\Delta f_{min}$.

Table \ref{TableMFSK}  contains the values of transmission rate corresponding to Fig. \ref{f4}.

\begin{table} [h]

    \caption[]{The transmission rate in bit/sec as a function of distance for un-coded non-coherent MFSK, transmitter $EIRP=10^{12}W$ and Arecibo-like receiver antenna, $SEFD=3J$, $BER=10^{-5}$. }

 \centering
    \label{TableMFSK}
   \begin{tabular}{c|c|c|c|c|c}
    \hline

     \multicolumn{6}{c}{\hspace{1.0cm} Distance, light years}\\
     \hline
    MFSK  & 10 & 50 &  100 & 500 &  1000 \\
     \hline
   M=2 & 27.2 & 1.09 &  0.27 & 0.01 &  $2.7 \cdot 10^{-3}$ \\
   M=4 & 51.2 & 2.05 &  0.51 & 0.02 &  $5.1 \cdot 10^{-3}$ \\
   M=8 & 72.5 & 2.9 & 0.725 & 0.03 & $7.0 \cdot 10^{-3}$ \\
   M=16 & 91.6 & 3.66 & 0.92 & 0.037 & $9.0 \cdot 10^{-3}$ \\
   M=32 & 108.8 & 4.35 & 1.09 & 0.044 & 0.011 \\
   M=64 & 124 & 5.0 & 1.24 & 0.05 & 0.012 \\
   \end{tabular}

\end{table}

\newpage

\section{Pattern detection}

\begin{figure}[h]
\epsfig{figure=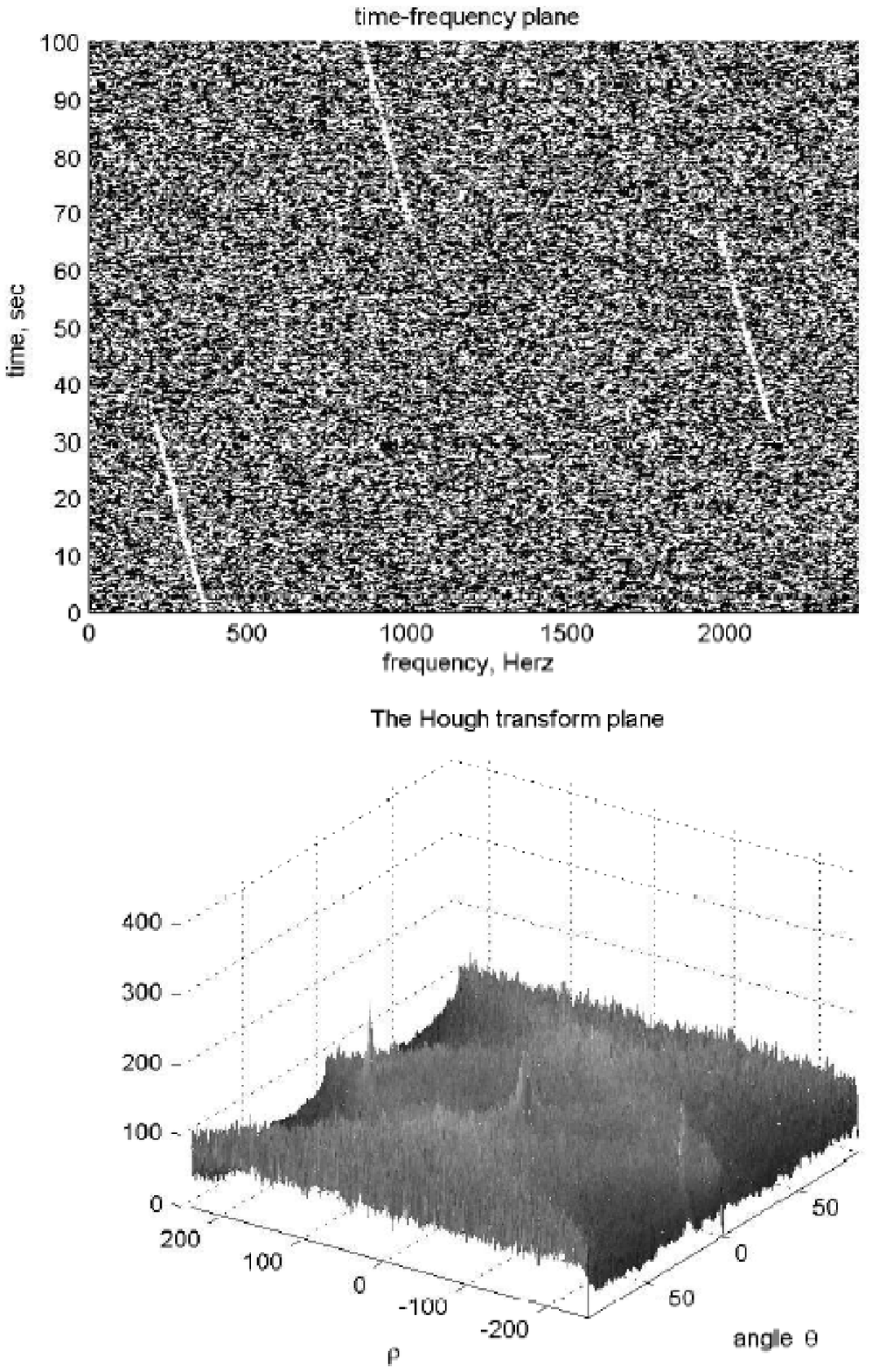,height=17cm,width=12.0cm}
\caption{Upper panel: three narrow-band impulses in the time-frequency plane imitate FSK signals. The lower panel: after the Hough transform, three peaks correspond to the three lines in the upper panel. The angle coordinate $\theta$ corresponding to the frequency drift is the same for all three peaks. The $\rho$-coordinates depict the difference in the carrier frequencies. }
\label{f7}
\end{figure}

The estimates of transmission rate given in the previous section support the presumption that it is  energetically advantageous to use MFSK, i.e.,  detection attempts can be focused on the search for narrow-band pulses scattered at  different frequencies. The duration of the pulses can be of the order of $20$ to $100s$ ($\Delta f_{min}=0.05$ to $0.01Hz$) or longer (if  reconciled with the necessity of long intervals of averaging,\citep{petr}). These signals are detectable on the time-frequency plane as random, non-intersecting parallel tracks. The slope of the tracks depends upon the Doppler frequency drift which might   not be compensated for at either transmitter or receiver side.  An artificial additional slope can  even be deliberately introduced by ETI to emphasize its artificial origin. Computer simulation was made to provide an example of such a pattern which is shown in the upper panel of Fig. \ref{f7}.

One of the possible methods of detection of such a pattern is to apply the Hough transform (HT) which is widely used in image processing for line detection \citep{duda}.  Details of HT and examples of computer simulation are given in Appendix B.  Here the detection of straight line tracks in the time-frequency plane with the help of HT is demonstrated.

The lower panel of Fig. \ref{f7} shows the result of HT as applied to the binary version (after thresholding) of the image in the upper panel. The  three peaks correspond to the three lines in the upper panel. The angle coordinate $\theta$ corresponding to the frequency drift (slope) is the same for all three peaks. The $\rho$-coordinates depict the difference in  carrier frequencies. No {\it a priori} information was used for detection: neither for the positions or durations of pulses or the slope of the lines.

When the track of a pulse in the time-frequency plane plane cannot be approximated by a straight line, HT can be modified to detect a second order curve. There are many versions of HT adapted to the detection of particular curves. The example in Appendix B (Fig. \ref{f10}) demonstrates the usage of HT for de-dispersion in the case of interstellar dispersion.

 An additional useful property of the HT is  the accumulating of  a number of  points belonging to the same  line, which is, in essence,  averaging.
Having the track of a pulse somewhere on the time-frequency plane  it is possible  also to detect  a weak pulse using this averaging property of HT, \citep{fridman}.

\newpage

\section{Conclusions}

1. Frequency-shift keying can be a method of modulation used by  ETI.  For long distances the low bit rate ($<0.01 bit/sec$) requires a long time on target to accumulate the necessary energy for the receiver. Modern SETI spectrum analyzers are well suited to searching for these types of signals.

2. Scattering and multi-path caused by  the  interstellar medium  broadens the sinusoidal  signal up to the bandwidth $df_{min} = 0.01$ to $0.05 Hz$. This means that  averaging of  CW for time intervals $>1/df_{min}$ sec must be performed in the bandwidth $\approx df_{min}$ which makes the transmission rate substantially slower.  Pulse signals with the same energy  but  shorter than $1/df_{min}$ (20 to 100 seconds) avoid this obstacle, because transmission of information by narrow-band pulses with the same average energy budget provides the same transmission bit rate. This can therefore be one of the  reasons to search for the narrow-band pulse-like signals from ETI.

3. Transmission power  of $10^{12}W$, which is equal to $10^{-5}$ of the Solar  radiation power coming to the Earth's surface,  is enough to receive  digital information with antennas such as  Arecibo and SKA with a transmission rate $RW \approx 10^{-2}$ bit/s at  a distance of 1000ly and an omni-directional transmitting antenna.  A transmission rate equal to $10^{-2}$ bit/s allows transmission of the full code of human genome ($1.2 \cdot 10^{10}$ bits) from a distance of 1000ly
in $1.2\cdot 10^{12}s=1.2\cdot 10^{5}years$, significantly less than the time taken by  natural evolution on Earth: $\approx 10^{9}years$.

If an ETI  were able to dispose of $n$ times larger transmission power, the transmission rate will proportionally increase.

4. The Hough transform can be a useful tool for detecting patterns in the time-frequency plane where there is  uncertainty about the  duration and frequency drift of narrow-band pulses.

\appendix

\section[]{Formulas for the  bit-error-rate}
1.
The probability of bit error for binary phase-shift keying (BPSK),\citep{Okunev}, BER can be estimated as follows:

\begin{equation}
P_{\rm BER,BPSK}=Q(\sqrt{2\times SNR_{\rm bit}}),
\end{equation}

\noindent where
\begin{equation}
Q(x)=\frac{1}{\sqrt{2\pi
}}\int\limits_{x}^{\infty}e^{\frac{-t^{2}}{2}}dt,
\end{equation}

\noindent $SNR_{\rm bit}=SNR_{\rm power}\times T_{bit}=\frac{SNR_{\rm
power}}{RW}$ is the signal-to-noise ratio per bit, $T_{bit}$ is the
duration of signal transmission per 1 bit and $RW=1/T_{bit}$ is the
transmission rate.

2. The probability of bit error for  coherent differential PSK (DPSK) is
\begin{equation}
P_{BER,DPSK}=2Q(\sqrt{2\times SNR_{bit}})[1-Q(\sqrt{2\times SNR_{bit}})]
\end{equation}

3. The probability of bit error for  non-coherent differential PSK is
\begin{equation}
P_{BER,noncorDPSK}=0.5\exp (-SNR_{bit})
\end{equation}
4. The probability of bit error for M-ary PSK
\begin{equation}
P_{BER,MPSK}=\frac{1}{k}2Q(\sqrt{2kSNR_{bit}}\sin \frac{\pi }{M})
\end{equation}

5. The probability of bit error for the {\it coherent} processing of
$M=2^{k}$ orthogonal signals (MFSK) can be calculated using the following
expression (\citep{proakis}, p. 259):
\begin{eqnarray}
P_{\rm BER,FSK\_C}=\frac{2^{k-1}}{2^{k}-1}\cdot
\frac{1}{\sqrt{2\pi }}\int_{-\infty }^{\infty
}[1-(\frac{1}{\sqrt{2\pi
}}\int\limits_{-\infty}^{y}e^{-\frac{t^{2}}{2}}dt)^{M-1}]\times\nonumber\\
\times exp[-(y-\sqrt{2\cdot SNR_{\rm symbol,FSK}})^{2}/2]dy, &  &
\end{eqnarray}
where $SNR_{\rm symbol,FSK}=SNR_{\rm power}\times T$, each signal
 transmits $k$ bits simultaneously.

 6. The probability of bit error for the {\it noncoherent} processing
of $M=2^{k}$ orthogonal signals is calculated using the following
expression (\citep{proakis}, p. 310):
\begin{eqnarray}
P_{\rm
BER,FSK\_NC}=\frac{2^{k-1}}{2^{k}-1}\sum_{n=1}^{M-1}(-1)^{n+1}(_{n}^{M-1})\times\nonumber\\
\frac{1}{n+1}\exp[-nkSNR_{\rm symbol,FSK}/(n+1)].
\end{eqnarray}

\section[]{The Hough Transform}
The Hough Transform (HT) is a useful algorithm for  the detection  of straight lines in  binary images  when the amplitudes of pixels are equal to two numbers, for example, 1 or 0.
Each straight line $y=ax+b$ in a plane can be parameterized by the angle $\theta$ of its normal to the horizontal axis and its distance $\rho$ from the origin of coordinates. The equation of a line is
\begin{equation}
y=-x\frac{\cos (\theta )}{\sin (\theta )}+\frac{\rho }{\sin (\theta )}.
\end{equation}
This equation can be rewritten for  $\rho$ as a function of $(x,y)$:
\begin{equation}
\rho=xcos(\theta)+ysin(\theta).
\end{equation}
A line can
then be transformed into a single point in the parameter space $(\rho, \theta)$ which  is
called the Hough space. For any pixel in the image with  position $(x, y)$, an infinite number of lines can go
through that single pixel. By using equation (B.2) all  pixels belonging to the line can be transformed into the Hough space.
A pixel is transformed into a sinusoidal curve that is unique for this pixel. Doing the same transformation for another pixel gives
another curve that intersects the first curve at one point in the Hough space. This point represents the straight
line  in the image space that goes through both pixels. This operation is repeated for all  pixels of the image.
The pixels belonging to the same straight line have the same point of intersection in the Hough space.

 For each point in the Hough space the HT program assigns a  counter which accumulates the number of these intersections. Therefore, if there is a straight line with the parameters $(\rho_{i}, \theta_{j})$ consisting of $n$ pixels the counter corresponding to the point $(i,j)$ in the Hough space will contain number $n$.
 This is correct only for the ideal case when there is no noise in the image. In the case of a noisy image the situation is as follows.

 Let the time-frequency  plane be the image on which the search for signal patterns is performed. Each line of pixels along the time axis represents  ``intensity'' samples at the i-th output of the spectrum analyzer's  filter bank.\\
 In the absence of  a signal pattern, these samples  are random numbers  - often with a  Gaussian pdf due to preliminary averaging. A binary image is created  using the threshold $thr$ equal to
 $thr \approx \sigma+m$
  where $\sigma$ is the {\it rms} of the noise samples and $m$ is the mean value. If the amplitude of a sample is less than $thr$ it is converted to 0, otherwise it is equal to 1. The binary image is covered with the random 0 and 1.

\begin{figure}[h]
\epsfig{figure=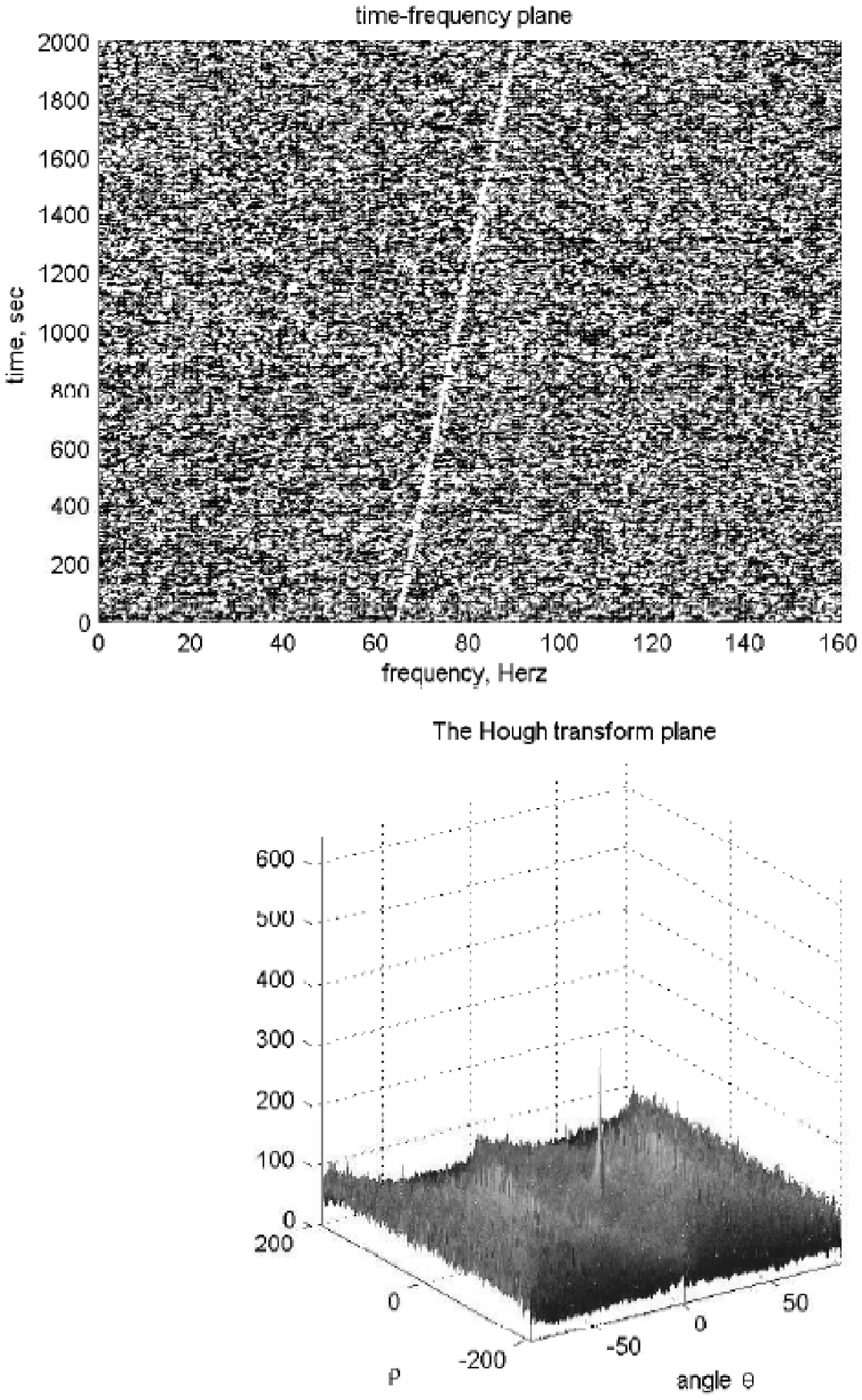,height=17.0cm,width=12.0cm}
\caption{The upper panel - time-frequency plane with the drifted narrow-band signal plus Gaussian noise, the lower panel-the Hough transform. The peak corresponds to $F_{0}=64.4 Hz$ and frequency drift slope $kf_{t}=0.0135 Hz/s$.}
\label{f8}
\end{figure}

  To better understand  the HT  a computer simulation example is given here. Let the  $400 \times 400$ image in Fig. \ref {f8}, upper panel, represent the ``noisy'' time-frequency  plane. Each horizontal line consists of $N=400$ samples of  the total power outputs of  the filter bank assigned to the frequencies from 0 to 160Hz. Each sample corresponds to the integration time $2000/N=5s$. The narrow-band signal is the input of the filter bank: with a frequency drift and the noise -  $x(t)=Scos[2\pi(F_{0}+kf \cdot t)t]+n(t)$,  the noise $n(t)$ is represented by  random numbers with the normal distribution $\cal{N}$(0,1), $S=0.2$. The lower panel of Fig. \ref {f8} shows the HT of the binary $400 \times 400$ image obtained from the image in the upper panel after  thresholding equal to 0.9 of the noise {\em rms}. The origin of coordinates is in the centre of the image. The peak due to the signal track  has coordinates $\rho=-6$ (in pixels) and $\theta=-9.45^{\circ}$ which correspond to the $F_{0}=64.4Hz$ and $kf=0.0135Hz/s$.

\begin{figure}[h]
\epsfig{figure=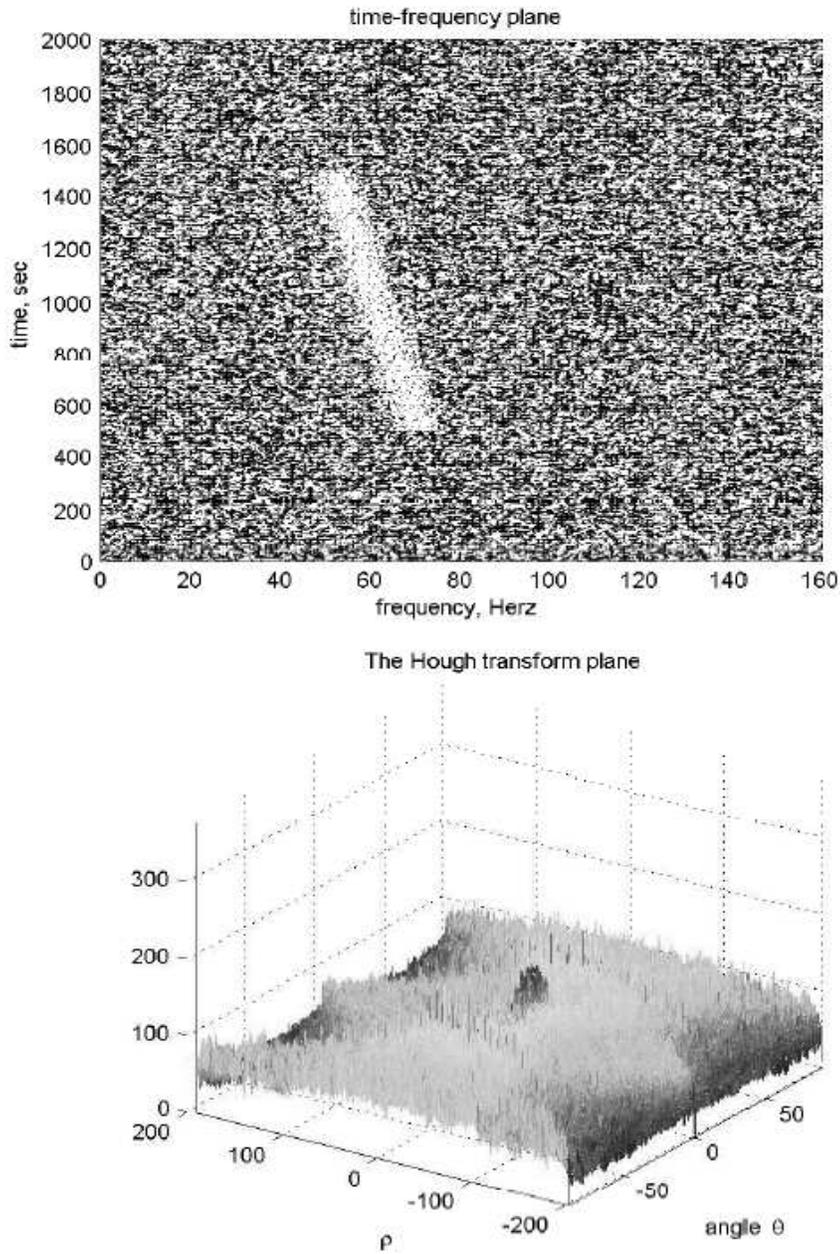,height=17.0cm,width=12.0cm}
\caption{The upper panel similar to Fig. \ref{f8} signal track in the time-frequency plane  but with the wider spectrum and different frequency drift;
the lower panel -  the Hough transform. The peak corresponds to $F_{0}=81 Hz$ and frequency drift slope $kf_{t}=-0.02 Hz/s$.}
\label{f9}
\end{figure}

  Another example in Fig. \ref {f9}, upper panel, illustrates a wide-band pulse signal in the time-frequency plane with a different frequency drift and the outputs of the same filter bank are simulated. The HT is shown in the lower panel of Fig. \ref {f9}. The peak corresponds to $F_{0}=81 Hz$ and frequency drift slope $kf_{t}=-0.02 Hz/s$.

\begin{figure}[h]
\epsfig{figure=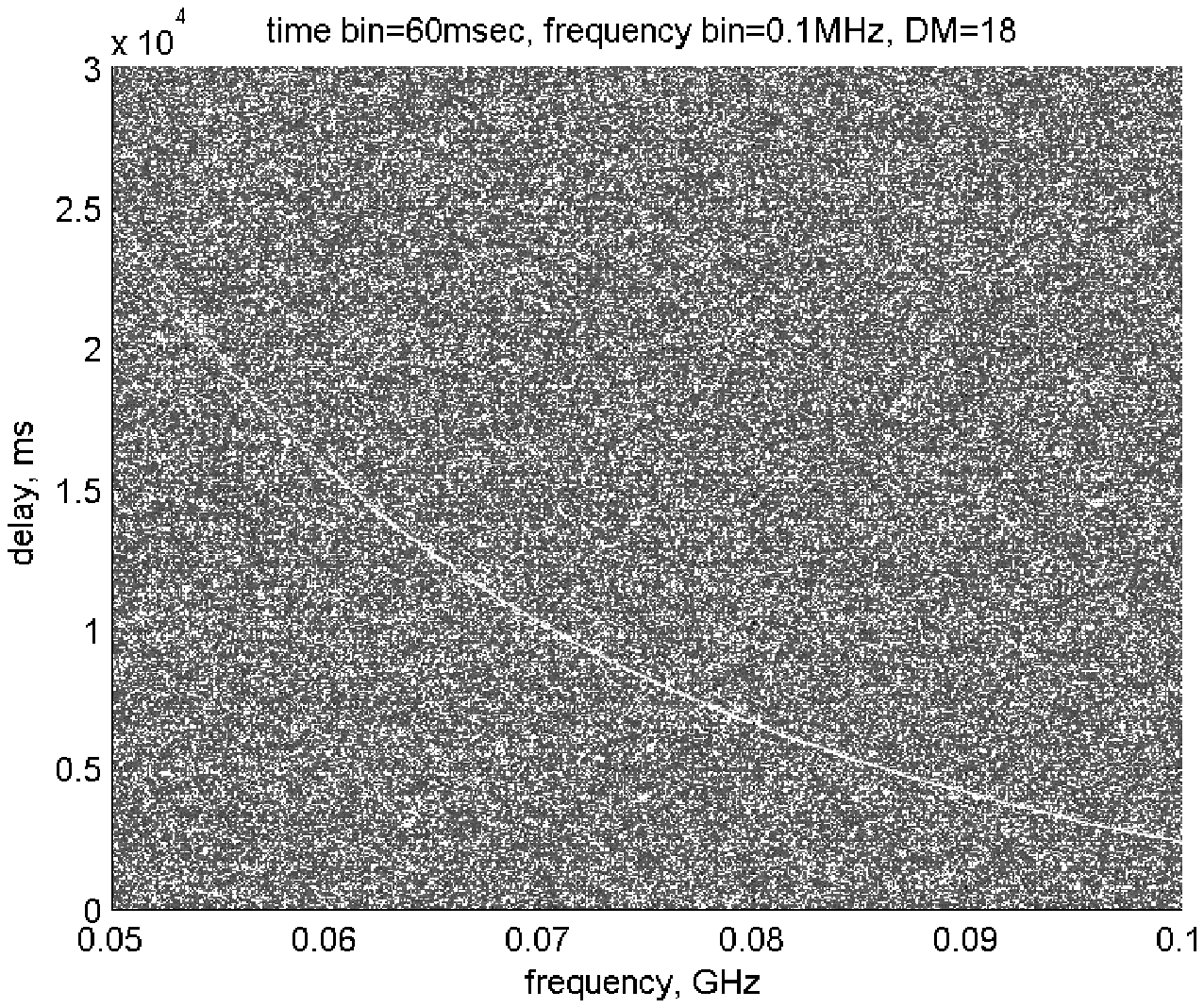,height=9.0cm,width=12.0cm}
\epsfig{figure=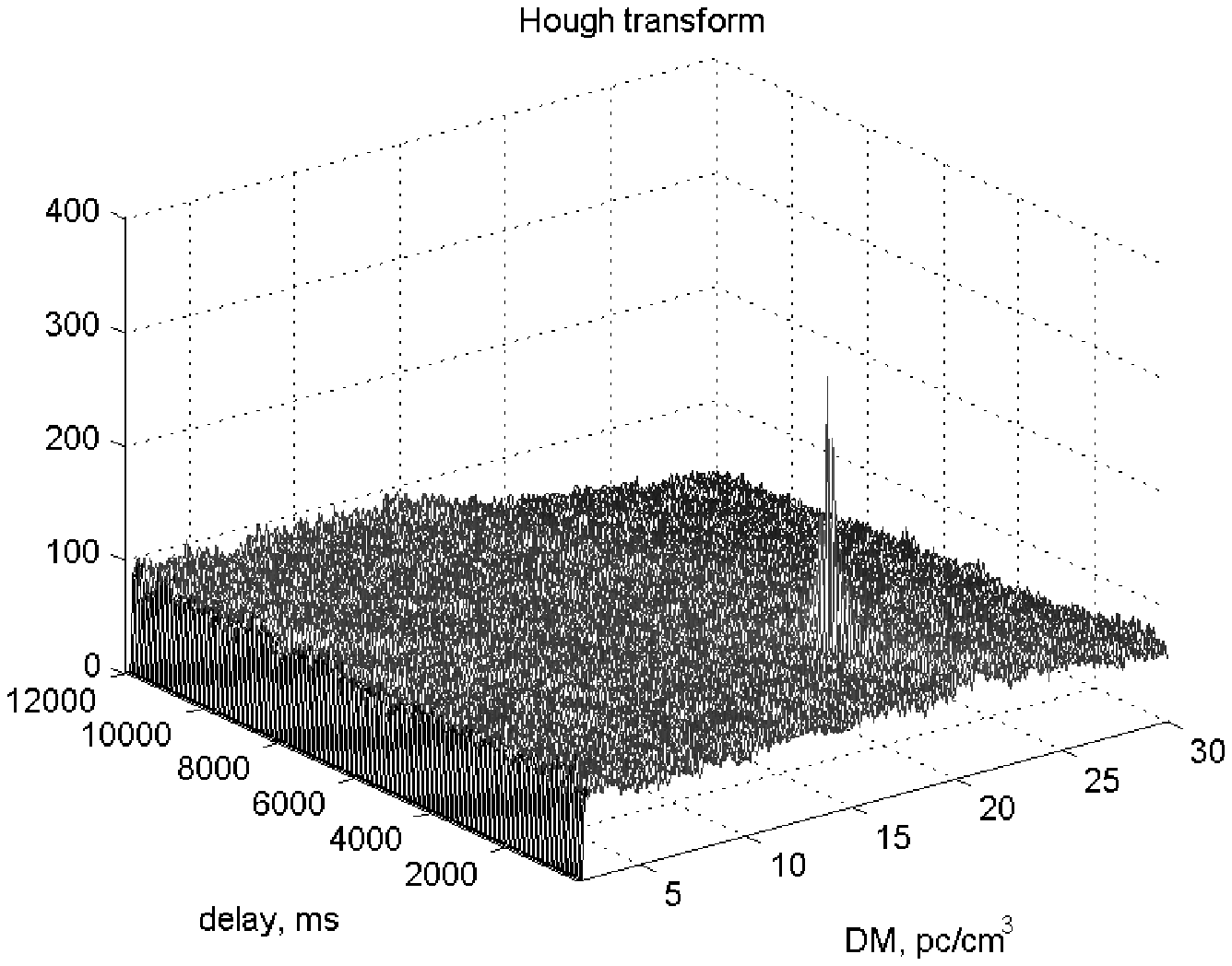,height=9.0cm,width=12.0cm}
\caption{ The upper panel - pulse track with dispersion (formula (B.2)) in the time-frequency plane. Two parameters: dispersion measure $DM$ and  delay at frequency 0.05GHz $dt_{0}$determine the position of the curve. The lower panel - the Hough transform. The coordinates of the peak correspond to $DM=18.04 pc/cm^{3}$ and  $dt_{0}=2400ms$.}
\label{f10}
\end{figure}

The last  example of the Hough transform  application is given in  Fig. \ref{f10} when the signal track on the time-frequency plane   is not linear: the interstellar dispersion  is calculated with the formula:
\begin{equation}
t_{delay}(f)=t_{0}(f_{1})+4.15DM[\frac{1}{f^{2}}-\frac{1}{f_{1}^{2}}],
\end{equation}
where
$t_{delay}-$ in msec, $f_{1}$ , $f$ in GHz, and $DM$ in $pc/cm^{3}$. Fig. \ref{f10}, upper panel, demonstrates a computer simulation of the pulse's spectrum with this kind of dispersion in the frequency range $0.05$ to $0.1GHz$. The parameter space for HT in this case is $[DM, t_{0}(f_{1})]$, i.e., the dispersion measure and the delay at the frequency $f_{1}=0.1GHz$. The coordinates of the peak in the HT plane in Fig. \ref{f10}, lower panel, correspond to $DM=18.04 pc/cm^{3}$ and  $dt_{0}=2400ms$.

\bibliographystyle{model1a-num-names}
\bibliography{<your-bib-database>}



\end{document}